\documentstyle[psfig]{mn}

\newcommand{\HI}{\mbox {\sc H\thinspace{i}}}
\newcommand{\HII}{\mbox {\sc H\thinspace{ii}}}
\newcommand{\Hline}[1]{\mbox{H{\footnotesize {#1}}}}
\newcommand{\Halpha}{\Hline{\mbox{$\alpha$}}}
\newcommand{\kms}{\mbox{km\thinspace s$^{-1}$}}
\newcommand{\Lya}{\mbox{Ly{\footnotesize$\alpha$}}}
\newcommand{\MHI}{\mbox{${\cal M}_{\rm HI}$}}
\newcommand{\Mlbstar}{\mbox{$({\cal M}/L_B)_\star$}}
\newcommand{\Mlbsun}{\mbox{$({\cal M}/L_B)_\odot$}}
\newcommand{\Msun}{\mbox{${\cal M}_\odot$}}
\newcommand{\Vrot}{\mbox{$V_{\rm rot}$}}
\newlength{\zwidth}
\newcommand{\zsp}{\mbox{\hspace{\zwidth}}}

\title[\HI\ and dark matter in NGC~1705]{\HI\ and dark matter in the 
windy starburst dwarf galaxy NGC~1705}
\author[G.R.\ Meurer, L.\ Staveley-Smith, and N.E.B.\
Killeen]{Gerhardt R.\ Meurer$^1$, Lister Staveley-Smith$^2$ and N.E.B.\
Killeen$^2$\\
$^1$The Johns Hopkins University, Department of Physics and Astronomy,
Baltimore, MD 21218, U.S.A.\\ 
$^2$Australia Telescope National Facility, 
CSIRO, P.O.\ Box 76, Epping, NSW 2121, Australia}

\date{Accepted 1998 June 12. Received 1998 June 1; in original form 1998 March 31}

\begin{document}

\maketitle

\begin{abstract}
We present 21cm \HI\ line observations of the blue compact dwarf galaxy
NGC~1705.  Previous optical observations show a strong outflow powered
by an ongoing starburst dominating the \HII\ morphology and kinematics.
In contrast, most of the \HI\ lies in a rotating disk.  An extraplanar
\HI\ spur accounts for $\sim 8$\%\ of the total \HI\ mass, and is
possibly associated with the \HII\ outflow.  The inferred mass loss rate
out of the galaxy's core is significant $\sim 0.2 - 2\, \Msun\, {\rm
yr^{-1}}$, but does not dominate the \HI\ dynamics.  Mass model fits to
the rotation curve show that the dark matter (DM) halo is dominant at nearly
all radii and has a central density $\rho_0 \approx 0.1\, \Msun\, {\rm
pc^{-3}}$: ten times higher than typically found in dwarf irregular
galaxies, but similar to the only other mass-modelled blue compact
dwarf, NGC~2915. This large difference strongly indicates that there is
little evolution between dwarf irregular and blue compact dwarf types.
Instead, dominant DM halos may regulate the morphology of dwarf galaxies
by setting the critical surface density for disk star formation.
Neither our data nor catalogue searches reveal any likely external
trigger to the starburst in NGC~1705.
\end{abstract}

\begin{keywords}
galaxies: individual: NGC~1705 -- galaxies: kinematics and
dynamics -- dark matter -- galaxies: ISM -- galaxies: starburst
\end{keywords}

\section{Introduction}\label{s:intro}

NGC~1705, at a distance $D = 6.2$ Mpc (for $H_0 = 75\, {\rm km\,
s^{-1}\, Mpc^{-1}}$; Meurer et al.\ 1995; hereafter M95)\nocite{m95}, is
a relatively nearby blue compact dwarf (BCD) galaxy, which in an earlier
paper (Meurer et al.\ 1992; hereafter Paper I\nocite{m92}), we showed
had several remarkable features.  These include: (1) a prominent $M_B =
-14.5$ ``super star cluster'', previously noted by Melnick, Moles,
Terlevich \shortcite{mmt85}, which we refer to as NGC1705-1 following
M95; (2) less luminous star clusters, \HII\ regions and a diffuse
distribution of massive stars which along with NGC1705-1 comprise an
intense ongoing starburst (see also M95); and (3) a spectacular galactic
wind (see also Marlowe et al.\ 1995\nocite{mhws95}), which is powered by
this starburst.  NGC~1705's explosive \Halpha\ morphology of multiple
loops and arcs extends out to the Holmberg radius $R_{\rm Ho} = 2.1$ kpc
and is accompanied by split emission lines.  The expulsive nature of the
flow is confirmed by UV absorption line kinematics (Heckman \&\
Leitherer, 1997\nocite{hl97}; Sahu \&\ Blades, 1997\nocite{sb97}).
Table~\ref{t:prop} lists some useful properties of NGC~1705 derived from
Paper I, M95, or this work.

\begin{table}
\caption{Derived and Adopted Parameters for NGC~1705\label{t:prop}}
\begin{tabular}{lcl}
Parameter & value & units \\ \hline
R.A.                     & ${\rm 04^h\, 54^m\, 13{\fs}50}$ & J2000 \\
Dec.                     & $-53^\circ 21'\, 39{\farcs}50$  & J2000 \\
$V_{\rm sys}({\rm Opt})$ & $628 \pm 9$                     & \kms \\
$D$                      & 6.2                             & Mpc \\
$R_{\rm Ho}$             & 2.1                             & kpc \\
$L_B$                    & $4.9 \times 10^8$               & $L_{B,\odot}$ \\
$L_{\rm FIR}/L_B$        & 0.23                            & \\
$V_{\rm sys}({\rm Dyn})$ & $640 \pm 15$                    & \kms \\
$\phi$(\HI\ major axis)  & 11                              & \degr \\
$i$(\HI)                 & 78                              & \degr \\
$R_{\rm HI}(5\times 10^{19}\, {\rm cm^{-2}\, contour})$ & $\approx 4.8$ & kpc \\
$V_\infty$               & 62.3                            & \kms \\ \hline
\end{tabular}
\end{table}

NGC~1705 and similar nearby BCDs (e.g.\ NGC~1569, NGC~5253, NGC~2915)
are ideal galaxies to study starbursts and their role in dwarf galaxy
evolution.  Star formation is an act of evolution, and the intense star
formation episode of a starburst, with its consequent gas consumption and
expulsion, may be the crucial transition event
in the evolution of a dwarf galaxy.  Starbursts in BCDs are also
easier to study than those in more luminous systems because BCDs tend
to have low dust content, and are more numerous (hence easier to find
locally) than high luminosity starburst hosts. Here we present a
detailed \HI\ study of NGC~1705 which addresses four important
questions concerning dwarf galaxy evolution and starbursts.

Firstly, what is the strength and fate of the winds often seen in BCDs
(Marlowe et al.\ 1995\nocite{mhws95}; 1997\nocite{mmhs97}; Hunter,
Hawley \&\ Gallagher 1993\nocite{hhg93})?  It has been proposed that
starburst driven winds may play a key role in enriching the inter
galactic medium (Heckman, Armus \&\ Miley 1990)\nocite{ham90}.  While
winds should escape more easily from low mass dwarf galaxies (e.g.\
Dekel \&\ Silk, 1986\nocite{ds86}), the extended dark matter halos
typically detected around galaxies increases their binding energies,
making it harder for outflows to escape.  The problem can be addressed
to the extent that the potential well depth can be determined, i.e.\
from a rotation curve analysis.  NGC~1705 is a particularly compelling
case for follow-up \HI\ study because its integrated \Halpha\ and \HI\
profiles have similar widths (Paper I).  The kinematics of the \Halpha\
are dominated by the outflow, hence this observation suggests that much
of the neutral ISM may also be entrained in the outflow.

Secondly, what triggers starbursts?  Is it an external perturbation,
such as an encounter, or due to some internal (secular) process such as
the formation of a bar (which of course can also be excited externally)?
While the most luminous starbursts occur almost exclusively in
interacting and merging systems \cite{v95}, BCDs tend to be fairly
isolated with respect to luminous galaxies \cite{sr94}, and hence an
external trigger is not always apparent.  Taylor (1997, and references
therein\nocite{t97}) looked for perturbing sources near BCDs and low surface
brightness dwarfs using \HI\ images from the VLA.  He found that the
incidence rate of HI companions near BCDs is 0.57 compared to 0.24 for
those near low surface brightness dwarfs, supporting the hypothesis that
the starbursts in BCDs are triggered by external interactions.
Nevertheless, their remains a high fraction ($\sim 40$\%) of BCDs that
show little or no signs of external triggers.

Thirdly, what are the evolutionary connections between different dwarf
galaxy morphologies?  There are two types of moderate to low surface
brightness dwarf galaxies.  Dwarf ellipticals (dEs) have very little or
no detectable ISM, smooth round isophotes, red colours, and often a
central nucleus.  Dwarf irregulars (dIs) have a high gas content,
irregular isophote shape, and blue colours.  A starburst may provide the
missing link between the two configurations.  That is, if a dI undergoes
a strong starburst, becoming a BCD, it can blow away its ISM in a
galactic wind and then fade to become a dE (e.g.\ Davies \&\ Phillips,
1988)\nocite{dp88}.

Finally, what is the relationship (if any) between dark matter (DM) and
starbursts?  The size and location of starbursts in infrared luminous
galaxies seems to be well governed by the global dynamics of the host;
starbursts tend to neatly fill the area covered by the rising portion of
the rotation curve, where rotational shear is low \cite{lh95}.  In dI
galaxies DM dominates often even interior to the rotation curve knee
(e.g.\ Carignan \&\ Beaulieu, 1989\nocite{cb89}; Lake, Schommer \&\ van
Gorkom 1990\nocite{lsv90}).  Could there be a difference in the (dark)
mass distribution of bursting and quiescent dwarfs?  Unfortunately,
while a great body of literature exists on the \HI\ and mass
distributions of dI galaxies (e.g. Begeman, Broeils, \&\ Sanders
1991\nocite{bbs91}), fewer BCDs have been imaged in \HI\ (e.g.\ Taylor
et al.\ 1995\nocite{tbgs95}), and only one has had its mass distribution
modelled.  That galaxy, NGC~2915, has a DM core density about 10 times
more dense than typically found in dIs (Meurer et al.\ 1996; hereafter,
M96)\nocite{m96}.  Is this a common property of BCDs?  If so what are
the implications?

Here we report on Australia Telescope Compact Array (ATCA) \HI\
synthesis observations of NGC~1705.  Preliminary results from this study
were presented by Meurer \shortcite{m94}.  In Section \ref{s:data} we
discuss the new 21cm data, and its reduction.  An overview of the radio
properties of NGC~1705 are presented in Section \ref{s:props}.  The
rotation curve of the galaxy is derived, and the mass distribution of
NGC~1705 is modelled in Section \ref{s:dyn}.  We discuss the
implications of our results in Section \ref{s:disc}.  Finally our
conclusion are presented in Section \ref{s:conc}.

\section{Data and reduction}\label{s:data}

NGC~1705 was observed with the ATCA over six runs, here labelled A -- F,
using different antenna configurations.  Five antennas were used in each
configuration.  The dual linear polarization AT receivers were employed
with the correlator set to 512 channels per polarization in each
baseline, and with each channel separated by 15.6 kHz.  The log of the
observations is given in Table~\ref{t:log}.  The observations were
comprised of sets were NGC~1705 was observed for about 40 minutes with
the phase centre set to 3$'$ north of the position of NGC1705-1 given in
Table~\ref{t:prop} followed by a 4 minute observation of the secondary
calibrator B0438-436 (${\rm RA = 4^h\, 40^m\, 17{\fs}180}$, ${\rm Dec =
-43\degr\, 33'\, 08{\farcs}60}$ (J2000), with VLBI positional accuracy
better than $0\farcs02$; Johnston et al.\ 1995)\nocite{j95}.  The
calibration assumes that the secondary calibrator is a point source with
constant flux density over the period of each observing run.  The
absolute flux calibration was set by observations of the primary
calibrator 1934-638\footnote{We adopt $F_\nu = 16.4$ Jy at 20cm for
1934-638 \cite{wo90}.  Reynolds \shortcite{r94} gives $F_\nu =
14.9$ Jy at 20cm for this source.}  which was observed at the beginning
or end of each observing run.  The one exception is run B, for which the
primary calibrator data was lost.  The flux density of B0438-436 is
reported in column 4 of Table~\ref{t:log} for the remaining five runs.
The mean 20 cm flux density of B0438-436 for configurations A,C,E,F is
$F_\nu = 5.41 \pm 0.05$ Jy which we adopt as the flux density of
B0438-436 during run B.  The flux density of B0438-436 during run D was
mildly discrepant from this mean.

\begin{table*}
\begin{minipage}{14cm}
\caption{ATCA observing log.\label{t:log}}
\begin{tabular}{ccccr}
Run & UT start     & Time on source & $F_\nu(B0438-436)$ & Baseline range \\
    & (dd/mm/yyyy) & (hours)        & (Jy)               & (m) \\ \hline
A & 17/11/1990 & ~3.5 & $5.31 \pm 0.13$ & 61 -- 2112 \\
B & 25/02/1991 & ~9.4 & \ldots          & 77 -- 1362 \\
C & 29/04/1991 & ~9.5 & $5.55 \pm 0.19$ & 337 -- 2924 \\
D & 25/05/1991 & 13.1 & $5.13 \pm 0.13$ & 31 -- 1286 \\
E & 17/07/1991 & 10.5 & $5.38 \pm 0.01$ & 31 -- ~459 \\
F & 01/12/1991 & ~8.9 & $5.40 \pm 0.02$ & 77 -- 1486 \\ \hline
\end{tabular}
\end{minipage}
\end{table*}
\begin{table*}
\begin{minipage}{15cm}
\caption{Dataset properties.\label{t:obs}}
\begin{tabular}{lcclcl}
 & \multicolumn{3}{c}{\HI\ Line images} & \multicolumn{2}{c}{continuum image} \\
Quantity & NA value & UN value & units & RUN value & units \\ \hline
image size  & $128 \times 128 \times 45$   & $128 \times 128 \times 37$ & 
pixels                                   & $1024 \times 1024$ & pixels \\
pixel size  & $10 \times 10 \times 6.6$    & $8 \times 8 \times 6.6$    & 
  ${\rm arcsec \times arcsec \times \kms}$ & $8 \times 8$       & ${\rm arcsec \times arcsec}$ \\
beam        & $37 \times 31$               & $25 \times 21$             & 
  ${\rm arcsec \times arcsec}$             & $34 \times 28$     & ${\rm arcsec \times arcsec}$ \\
noise/pixel & 1.3                          & 1.9                        & 
  mJy beam$^{-1}$                          & 0.25               & mJy beam$^{-1}$ \\
$T_b$ (max) & 19                           & 25                       & 
 K                                    & 80                 & K \\ \hline
\end{tabular}
\end{minipage}
\end{table*}

The data processing was very similar to that used by M96, to which the
reader is referred for details.  The steps included data editing,
calibration, combining the data from the two polarizations, continuum
subtraction (using 300 channels free of line emission), shifting of the
data to a common heliocentric rest frame, time averaging to one minute
per visibility, combining the individual data sets, and then imaging and
``CLEAN-ing'' the data (Schwab, 1984; see also Clark, 1980; H\"ogbom,
1974)\nocite{s84}\nocite{c80}\nocite{h74}. The resulting uniformly (UN)
weighted and naturally (NA) weighted data cubes were made by averaging
two spectral channels at a time during the imaging stage resulting in a
channel separation of 6.6 \kms.

\begin{figure*}
\begin{minipage}{14cm}
\centerline{\hbox{\psfig{figure=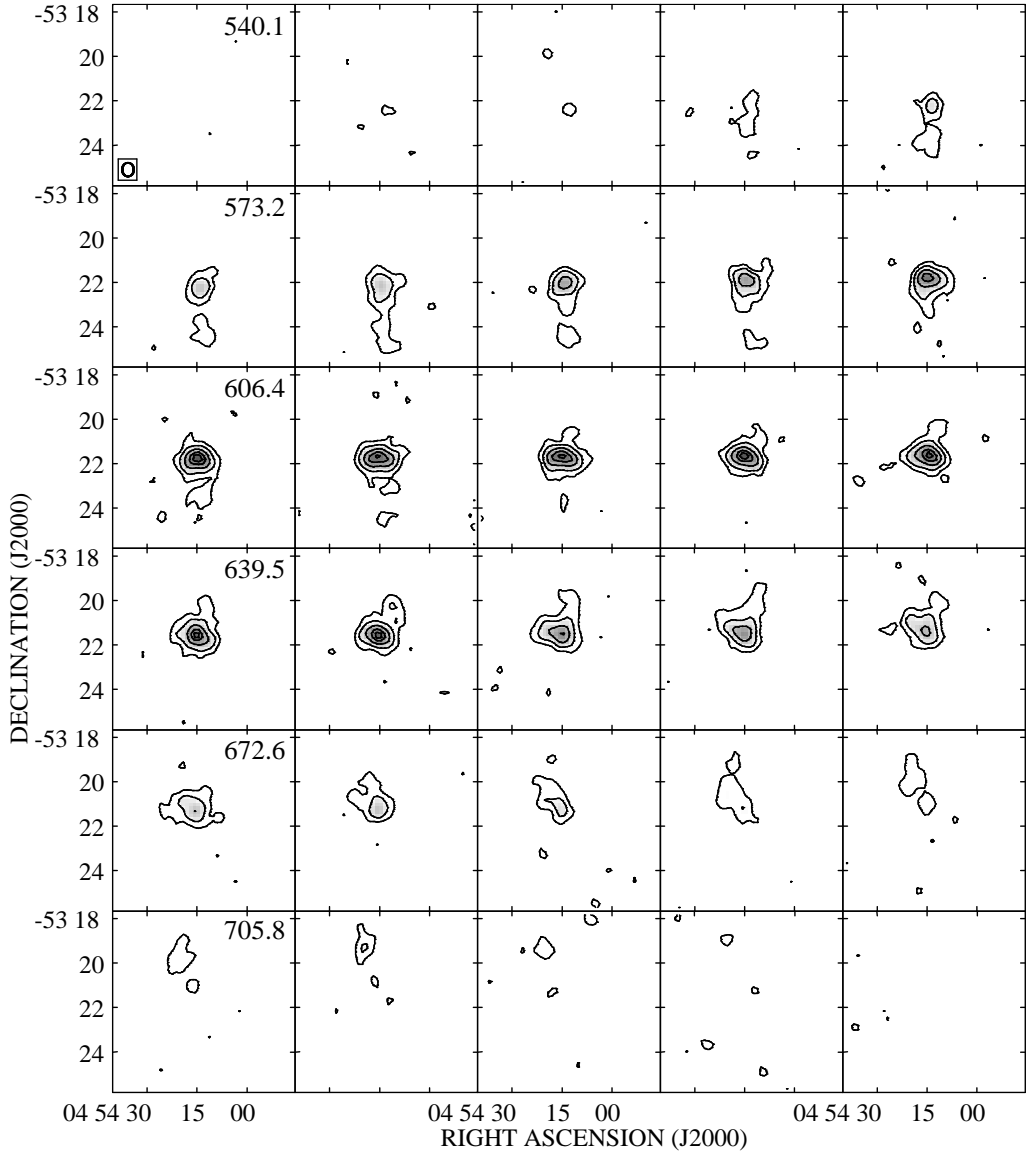,width=14cm}}}
\vspace{-1.5cm}
\caption{The continuum subtracted, cleaned NA data cube
showing planes of constant velocity over the $V_r$ range of strongest
\HI\ emission.  The convolving beam size is shown in the lower left
corner of the upper left panel.  The $V_r$ of every fifth plane in \kms\
is listed in its upper right corner.  Contours are drawn at 10, 25, 50,
75, and 90 percent of the peak flux density of 35.7 mJy beam$^{-1}$
($N_{\rm HI} = 2.27 \times 10^{20}\, {\rm cm}^{-2}$). \label{f:cube_na}}
\end{minipage}
\end{figure*}

A preliminary examination of the the cleaned NA continuum image showed
significant calibration errors.  Therefore we improved the calibration
by self-calibrating \cite{pr84} with this continuum image.  This
resulted in a 24\%\ improvement in the noise level of the natural (NA)
weighted continuum image to 0.31 mJy beam$^{-1}$, and a 7\%\ improvement
in the NA weighted \HI\ data cube to 1.25 mJy beam$^{-1}$. Planes of the
NA data cube containing \HI\ signal are shown in Fig.~\ref{f:cube_na}.
The final continuum image was made using ``robust'' uniform (RUN)
weighting \cite{b95}.  A summary of the properties of the data sets
analysed in this paper are presented in Table~\ref{t:obs}.

\section{21cm properties}\label{s:props}

\subsection{HI properties}

\begin{figure*}
\begin{minipage}{16.8cm}
\centerline{\hbox{\psfig{figure=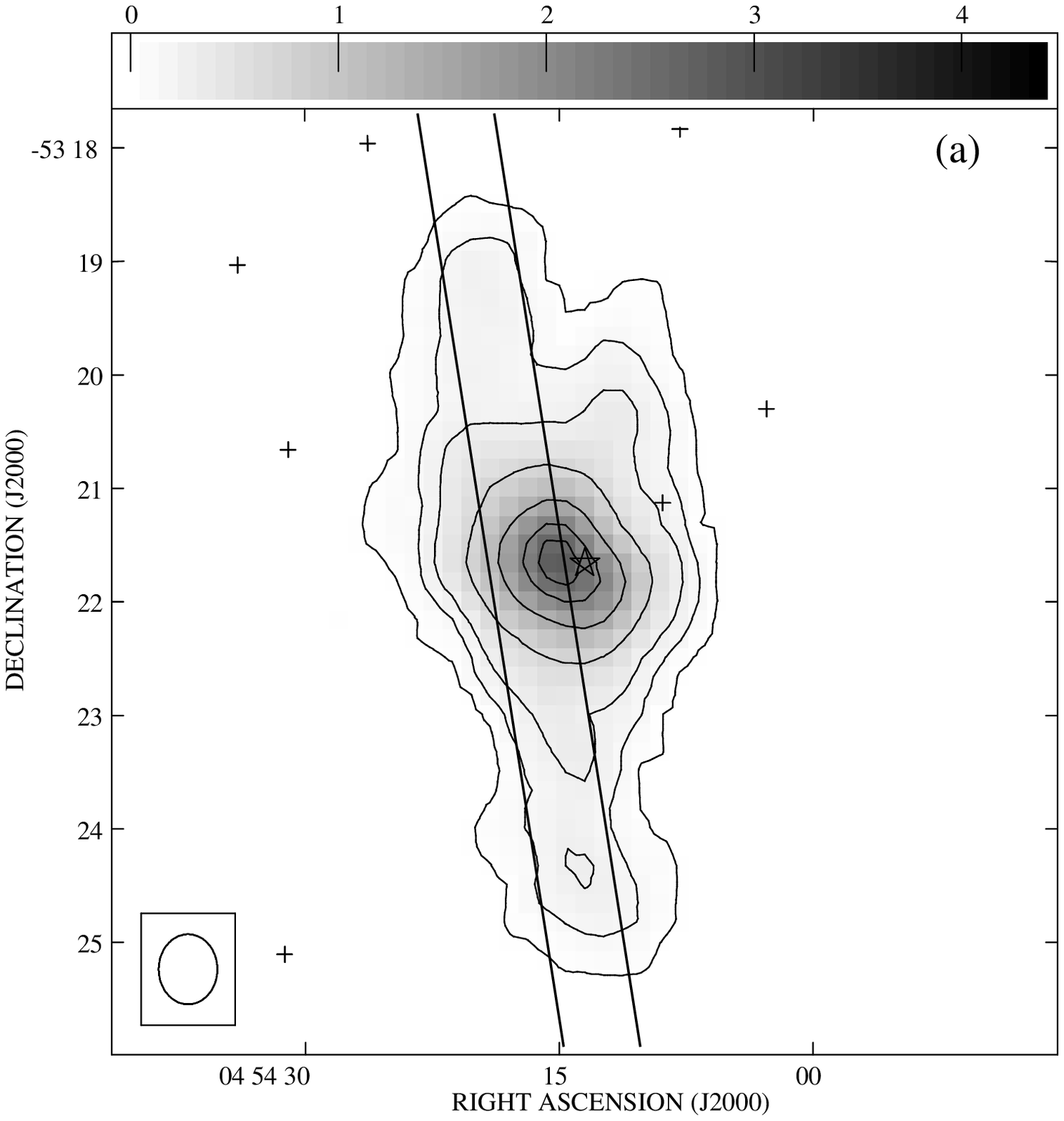,width=8.4cm}\psfig{figure=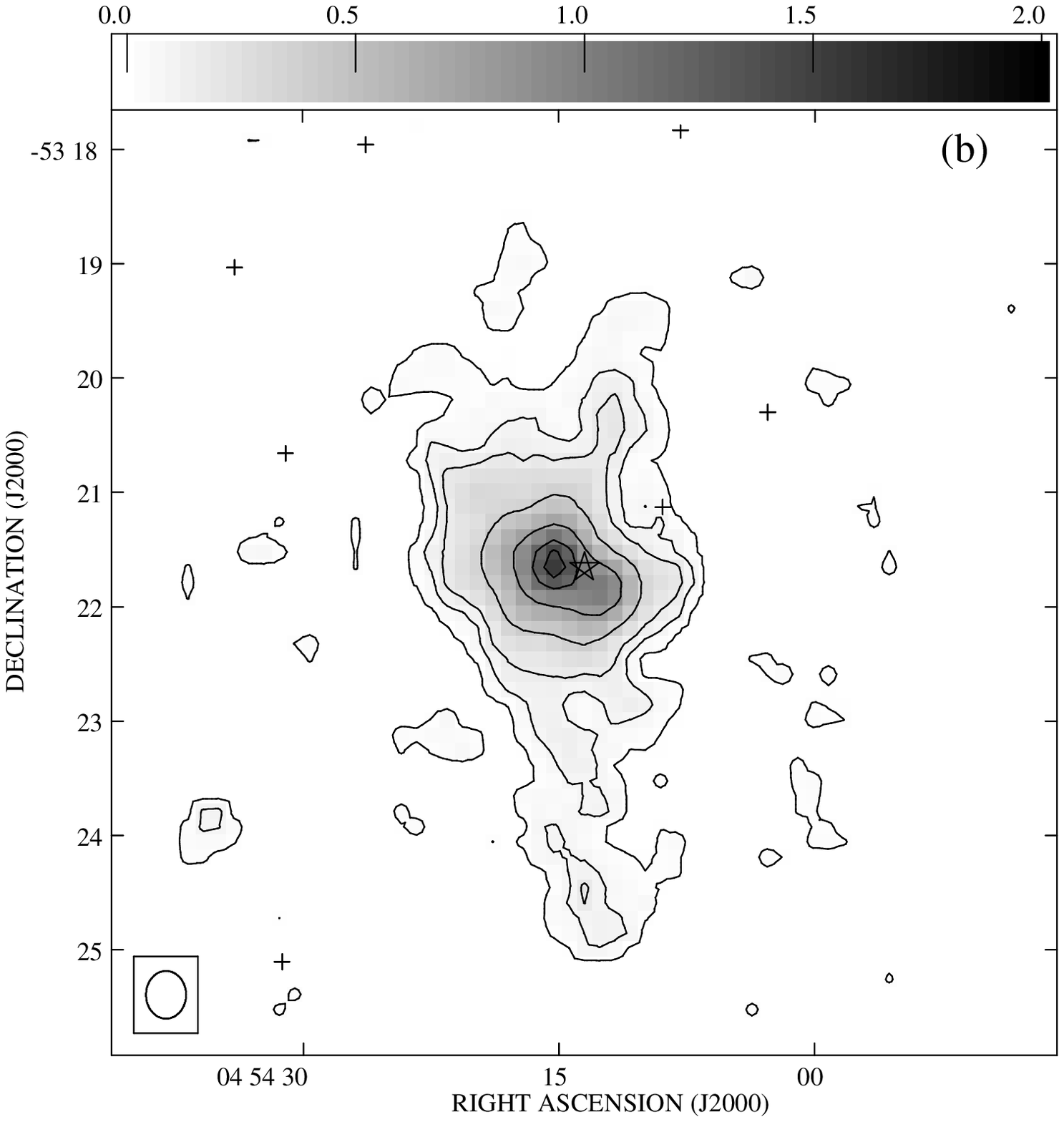,width=8.4cm}}}
\vspace{-1.cm}
\caption{Total \HI\ column density as determined from the zeroth moment
of the NA (panel a) and UN (panel b) \HI\ data cubes.  Contour levels
correspond to 1, 5, 10, 25, 50, 75, and 90 percent of the peak flux
density which in the NA map is 2800 Jy/Beam m/s ($N_{\rm HI} = 2.7\times
10^{21}\, {\rm cm^{-2}}$) and in the UN map is 1540 Jy/Beam m/s ($N_{\rm
HI} = 3.2\times 10^{21}\, {\rm cm^{-2}}$).  The two nearly vertical
lines in panel a delimit the slice plotted in Fig.~\ref{f:pv1_na}. In
this and other images, the beam size is shown at lower left, the
position of NGC1705-1 is indicated with a star, and field stars are
indicated with plus signs.\label{f:mom0} }
\end{minipage}
\end{figure*}

\begin{figure*}
\begin{minipage}{16.8cm}
\centerline{\hbox{\psfig{figure=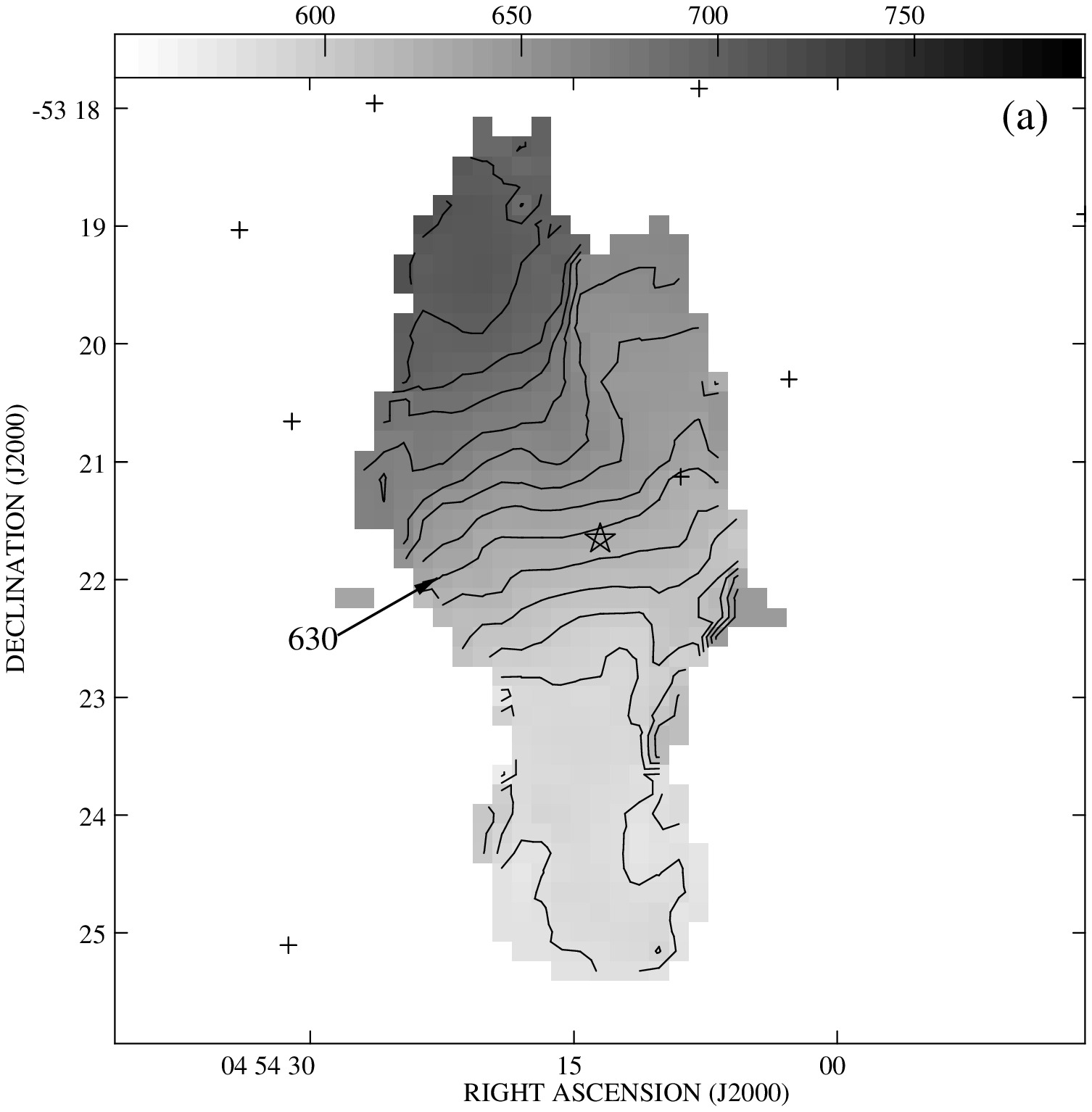,width=8.4cm}\psfig{figure=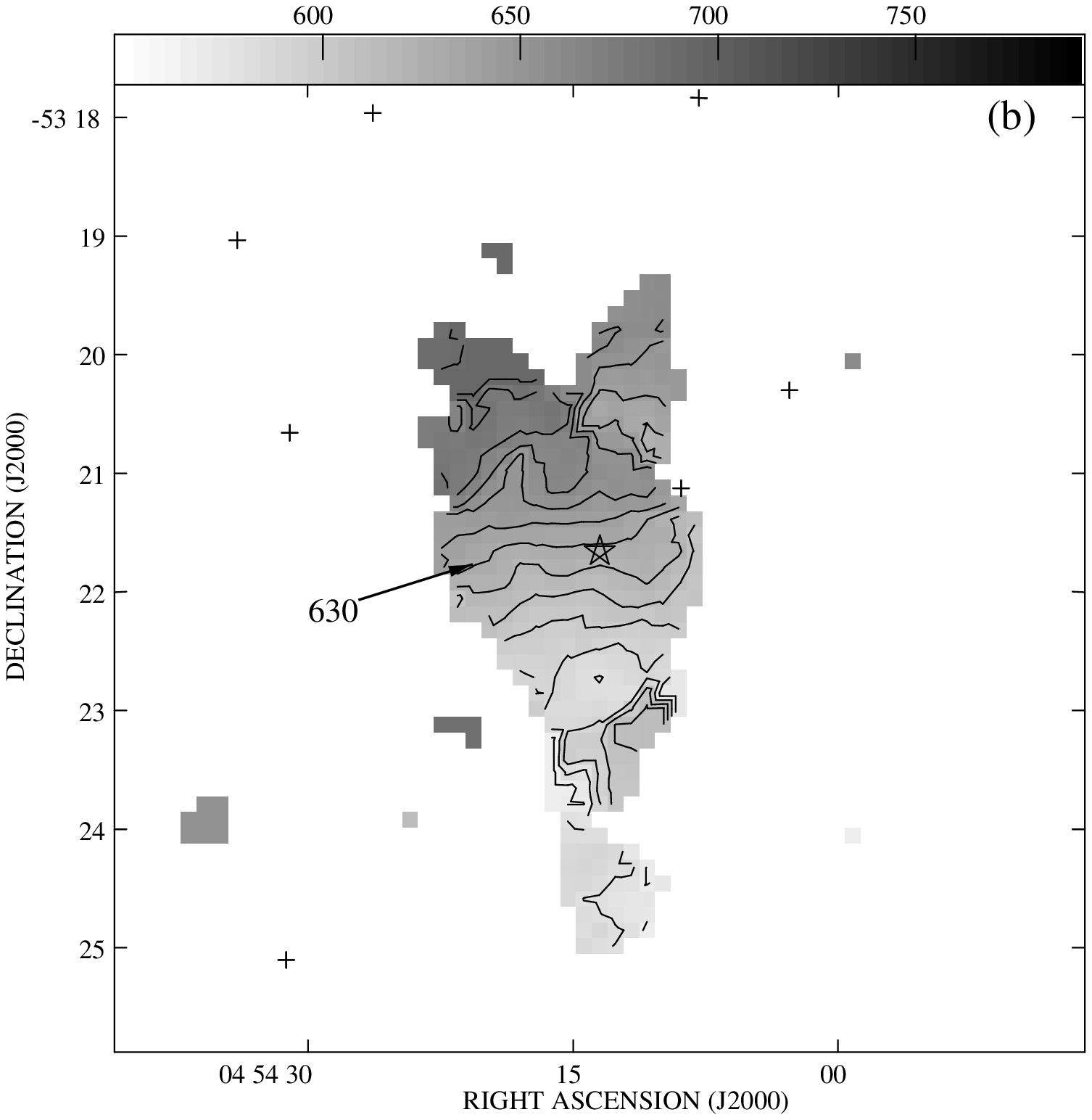,width=8.4cm}}}
\vspace{-1.cm}
\caption{Velocity field determined from the first moment of the NA
(panel a) and UN (panel b) data cubes.  Contours are drawn at 10 km
s$^{-1}$ intervals, with the $V_r = 630\, {\rm km\, s^{-1}}$ contour
indicated.\label{f:mom1}}
\end{minipage}
\end{figure*}

\begin{figure*}
\centerline{\hbox{\psfig{figure=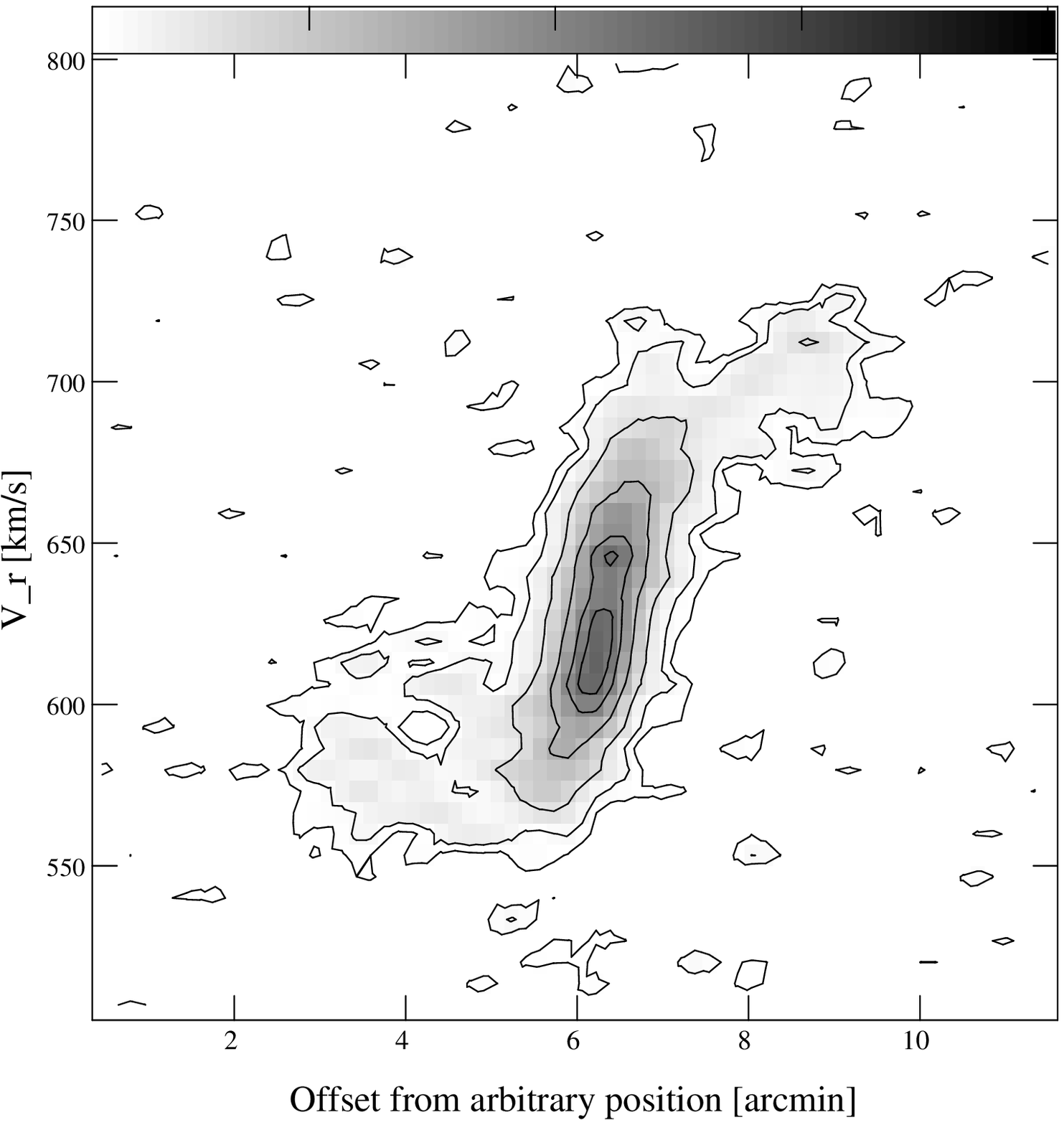,width=8.4cm}}}
\vspace{-0.5cm}
\caption{ Position - Velocity (PV) diagram along the \HI\
major axis.  This was made from a 40 arcsec  wide cut in the NA data cube
between the lines shown in Fig.~\ref{f:mom0}.
\label{f:pv1_na}}
\end{figure*}

The distribution and dynamics of the \HI\ can largely be surmised from
Figs.~\ref{f:mom0} - \ref{f:pv1_na}.  Figure~\ref{f:mom0} shows the
total \HI\ distribution of NGC~1705 in the NA and UN zeroth moment
maps. The \HI\ is strongly peaked near the position of NGC1705-1.  At UN
resolution the central \HI\ concentration starts to resolve into two
peaks separated by $\approx 30''$ (0.9 kpc) which straddle
NGC1705-1. The valley between the peaks is reminescent of the hole in
the \HI\ distribution around the brightest cluster in the similar galaxy
NGC~1569 \cite{iv90}.  This was attributed by Israel and van Driel to
supernovae evacuating a cavity in the ISM.  At lower surface
brightnesses the \HI\ distribution is elongated along a position angle
$\phi = 9\degr$ (measured relative from north towards the east), and
there is a strong velocity gradient along this axis, as shown in
Figure~\ref{f:mom1} which plots the NA and UN velocity fields obtained
from the first moment of the data cubes.  The overall impression is that
we are viewing a highly inclined disk.  This impression is supported by
Fig.~\ref{f:pv1_na} which shows a position-velocity (PV) cut (40 arcsec
wide) along the \HI\ major axis, here rotation is clearly evident.


The \HI\ peak(s), and NGC1705-1 are clearly displaced away from the
symmetry point of the PV cut (Fig.~\ref{f:pv1_na}), and these centres
are displaced above the mid-line of the \HI\ plane
(Fig.~\ref{f:mom0}). The offsets amount to 25 arcsec  radially and 27 arcsec 
out of the \HI\ plane. The orientation of the optical isophotes
suggests that NGC~1705 is a warped system with the centre being more
face-on than the extremities.  Section~\ref{s:dyn} discusses the
dynamics of the disk in more detail.

Also apparent in Fig.~\ref{f:mom0} is a spur of \HI\, at $\phi = -10\degr$
relative to the optical centre extending to a projected $R \approx 2.5'
= 4.5$ kpc. The velocity contours in Fig.~\ref{f:mom1} show that it is a
kinematically distinct structure.  We discuss possible origins of the
spur in Section \ref{s:disc}. In addition, the \HI\ line profiles are split
$\sim 130$ arcsec  S of the optical centre, as can be seen in
Fig.~\ref{f:pv1_na} (see also Fig.~\ref{f:lprofs} below), kinematically
suggesting the presence of an expanding bubble or a high velocity cloud.

\begin{figure*}
\centerline{\hbox{\psfig{figure=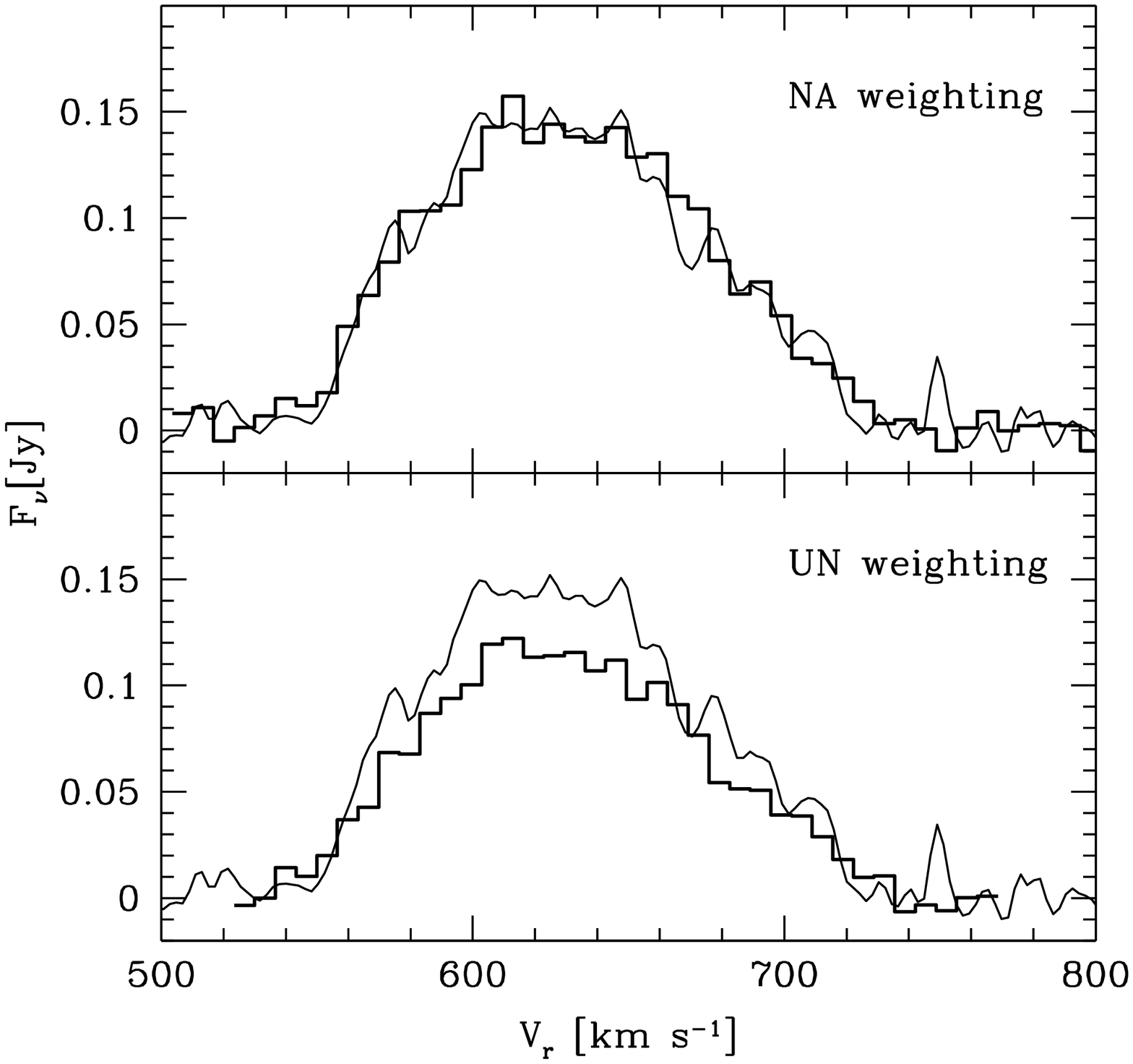,width=8.4cm}}}
\vspace{-0.5cm}
\caption{Globally integrated \HI\ velocity profiles from the NA and UN
data cubes are drawn as histogram style lines.  For comparison the
Parkes 64m spectrum (Paper I) is shown as the thin jagged
line.\label{f:globprof}}
\end{figure*}

The NA and UN integrated velocity profiles are shown in
Fig.~\ref{f:globprof} which shows that the UN data is missing flux
relative to the NA data (by $\approx 20$\%).  Properties of the data
measured from the global profiles are presented in
Table~\ref{t:glob}. The NA integrated flux agrees well with that
measured from Parkes single dish spectra $\int S d\nu = 16.6 \pm 0.2
{\rm Jy\, \kms}$ (Paper I).

\begin{table}
\caption{Measurement of the global \HI\ profile\label{t:glob}}
\begin{tabular}{lccl}
Quantity & NA value & UN value & units  \\ \hline
$\int S d\nu$          &  16.9   & 13.2  & Jy \kms \\
$M_{\rm HI}$           &  1.53   & 1.19  & $10^8$ \Msun \\
$V_{\rm sys}$          &  630.2  & 630.2 & \kms \\
$W_{50}$               &  119    & 112   & \kms \\
$W_{20}$               &  163    & 163   & \kms \\ \hline
\end{tabular}
\end{table}

\subsection{Continuum properties}

\begin{figure}
\centerline{\hbox{\psfig{figure=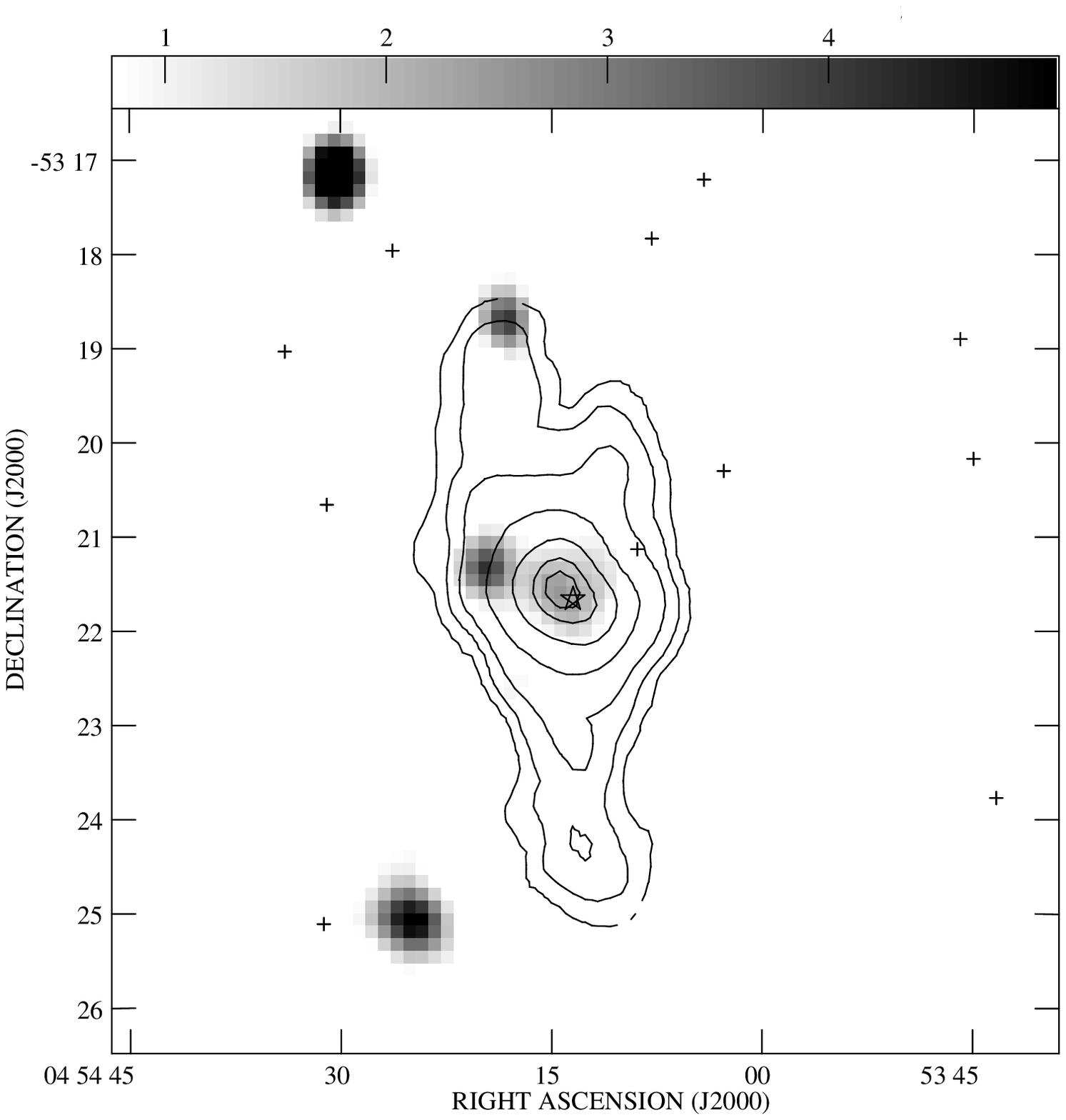,width=8.4cm}}}
\vspace*{-1.0cm}
\caption{Radio continuum image (1.4 GHz) in greyscale
overplotted with the \HI\ line image in contours.  The continuum flux
density in mJy/Beam is shown by the bar on top.  Contour levels
correspond to 2, 5, 10, 25, 50, 75, and 90 percent of the peak flux
density of 2800 Jy/Beam m/s.\label{f:cont}}
\end{figure}

Figure~\ref{f:cont} shows the continuum image with the NA zeroth moment
\HI\ isophotes overlayed.  Two sources are seen near the optical
position of NGC~1705.  Their flux densities and positions were measured
with a two component Gaussian fit.  The source located at $\Delta\alpha
= 55{\farcs}1 \pm 1{\farcs}5$, $\Delta\delta = 20{\farcs}9 \pm
1{\farcs}2$ with respect to NGC1705-1 is unresolved and has a flux
density $F_\nu = 4.2 \pm 0.4$ mJy.  While no optical sources are
coincident within the errors of this position, there is a faint point
like source about 13 arcsec  away from this position, i.e.\ within the
radius of the RUN beam.  The other source coincides well with the the
position and size of the galaxy in the optical.  It has $F_\nu = 8.0 \pm
0.4$ mJy, but appears relatively faint in Fig.~\ref{f:cont} because it
is resolved, having half power dimensions $56\pm 2'' \times 42\pm 4''$,
very similar to the dimensions in \Halpha\ and optical continuum.  Note
that the major axis position angle $\phi = -41\pm 11\degr$, is orthogonal
to the optical continuum $\phi = 50\degr$, and similar to the \Halpha\
outflow axis $\phi \approx -20\degr$.

The 1.4GHz continuum flux density is somewhat higher than expected by
the radio-far infrared (FIR) correlation.  Using the {\it IRAS\/} fluxes listed
in Paper I and following Devereaux \&\ Eales \shortcite{de89}, we find
$L_{\rm FIR} = 6.1\times 10^7 L_\odot$ while the continuum power is
$P_{\rm 1.4 GHz} = 3.5\times 10^{19}\, {\rm W\, Hz^{-1}}$.  For the
observed $L_{\rm FIR}$, the regression line of Devereaux \&\ Eales
predicts $P_{\rm 1.4 GHz} = 6.8\times 10^{18}\, {\rm W\, Hz^{-1}}$ a
factor of five lower than observed (the scatter in the correlation is
0.34 dex, a factor of 2.2).

This result may be due to the low dust content of NGC~1705.  M95 find
NGC~1705 to have the bluest ultraviolet (UV) colours and lowest FIR/UV
flux ratio of any of the galaxies in their sample except for IZw18,
indicating that NGC~1705 has a very low dust content (see their Tables
4, 8, and Fig.~6).  Since it is radiatively heated dust which provides
the FIR flux, the low $L_{\rm FIR}$ results from the low dust content,
rather than $P_{\rm 1.4 GHz}$ being abnormally high.

\section{Dynamics}\label{s:dyn}

\subsection{Extraction of the rotation curve}\label{ss:rc}

\begin{figure}
\centerline{\hbox{\psfig{figure=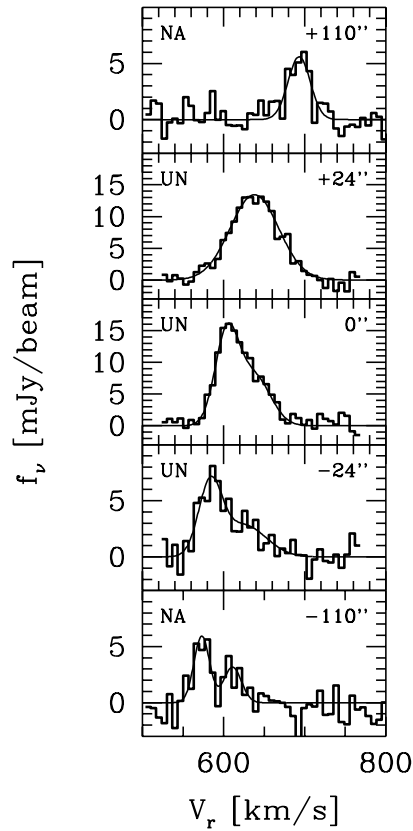,height=12cm}}}
\caption{Example \HI\ line profiles (histogram style line) and their
Gaussian fits (smooth line).  The profiles were extracted from the slice
along the \HI\ plane indicated in Fig.~\ref{f:mom0}. Each panel shows
the radial offset from the projected position of NGC1705-1 in the upper
right (+ indicates towards the NNE), and whether the profile was
extracted from the UN or NA slice in the upper right.  The profile in
the second panel from the top corresponds closely to the dynamical
center, while the bottom most profile was extracted within the kinematic
bubble apparent in Fig.~\ref{f:pv1_na}.
\label{f:lprofs}}
\end{figure}

The rotation curve was determined from 40 arcsec  wide slices of the UN and
NA data cubes along the \HI\ plane (e.g. Fig.~\ref{f:pv1_na}).  This
width was selected to match the disk size, and yet avoid the spur
region.  The line profiles in these images were fitted with Gaussians to
find $V_r$.  Example profiles and their fits are shown in
Fig.~\ref{f:lprofs}.  In cases where the line profiles were multiple or
asymmetric (as expected for the inner regions of a highly inclined disk)
the component with $V_r$ farthest from $V_{\rm sys}$ was used to
estimate the rotational amplitude. 

First the dynamical centre and its corresponding systemic velocity
$V_{\rm sys}({\rm dyn})$ were determined by finding the symmetry point of
the NA $V_r$ measurements.  This yields  $V_{\rm sys}({\rm dyn}) = 640 \pm
15$ \kms\ and the rotation axis displaced $25 \pm 13''$ ($740 \pm 390$
pc) NNE (along the principal \HI\ axis) from the projected position of
the NGC1705-1 on the \HI\ disk.  Note that the lower values of $V_{\rm
sys}$ determined optically (Paper I) and from the integrated \HI\
profile, 628, 620 \kms\ respectively, are consistent with this
displacement between the optical and dynamical centres.

The rotation curve was then extracted by taking $R$ to be the radial offset
from the dynamical centre, and taking $V_{\rm rot}\sin i = V_r - V_{\rm sys}({\rm
dyn})$.  The column by column $V_{\rm rot}\sin i$ measurements from the 2D
slices are averaged within bins equal to the (approximate) relevant beam size
yielding the adopted $V_{\rm rot}\sin i$.  Our adopted rotation curve uses
the UN measurements for $R \leq 75''$ and the NA measurements for $R > 75''$.

\begin{figure}
\centerline{\hbox{\psfig{figure=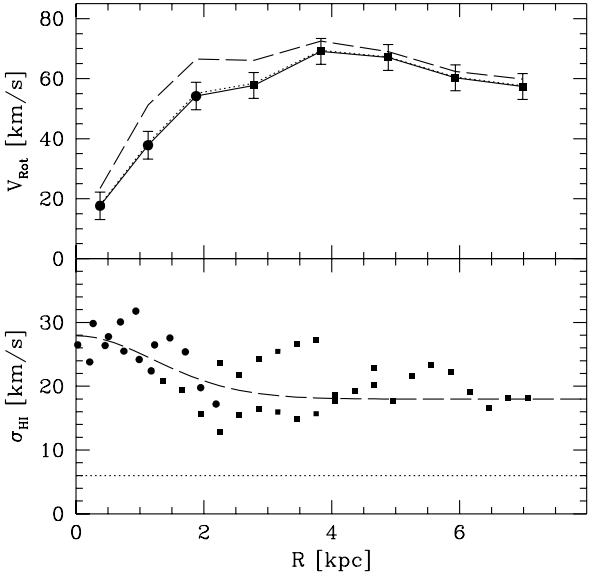,width=8.4cm}}}
\caption{The points in the top panel, connected with the solid line,
show the derived rotation curve.  The bottom panel shows the raw line of
sight \HI\ velocity dispersion, $\sigma_{\rm HI}$.  In both panels
measurements of the UN and NA data are plotted as circles and squares
respectively. The data in the lower panel are the individual
measurements from the PV slices (e.g.\ Fig~\ref{f:pv1_na}). Receding and
approaching sides are plotted separately.  The dotted and dashed lines
portray two plausible models for the face-on $\sigma_{\rm HI}$ profile,
corresponding to minimum and maximum pressure support respectively.  The
pressure support corrected circular velocities $V_c$ are plotted with
the corresponding line styles in the upper panel.  \label{f:rc}}
\end{figure}

A dynamical inclination can not be determined directly from the velocity
field since the velocity contours do not splay out along the minor axis
into the typical spider pattern of a moderately inclined disk.  This in
part is due to the contamination of the spur, but is also indicative of
a highly inclined disk.  Instead $i= 78\degr$ was estimated from the
shape of the outer \HI\ isophotes (excluding the ``spur'' region),
assuming an intrinsic minor to major axis ratio $c/a = 0.22$
\cite{bgp83}.  If the true inclination is lower in the centre as implied
by the optical isophotes, then the central two $V_{\rm rot}$ values may
be higher by a factor $\la 1.5$.  The adopted rotation curve is
given in Table~\ref{t:rc}, and plotted in the upper panel of
Fig.~\ref{f:rc}.  The average errors were estimated separately for the
UN and NA data as half the $V_{\rm rot}$ difference from opposite sides
of the galaxy, and are listed in Table~\ref{t:rc}.

\begin{table}
\caption{Adopted rotation curve.\label{t:rc}}
\begin{tabular}{rr}
R & $V_{\rm rot}$ \\
(arcsec) & \kms \\ \hline
 12.5 &  $17.6 \pm 4.6$ \\
 37.5 &  $37.8 \pm 4.6$ \\
 62.5 &  $54.2 \pm 4.6$ \\
 92.5 &  $57.7 \pm 4.3$ \\
127.5 &  $69.1 \pm 4.3$ \\
162.5 &  $67.0 \pm 4.3$ \\
197.5 &  $60.3 \pm 4.3$ \\
232.5 &  $57.4 \pm 4.3$ \\ \hline
\end{tabular}
\end{table}

\subsection{Surface density profiles\label{ss:prof}}

\begin{figure}
\centerline{\hbox{\psfig{figure=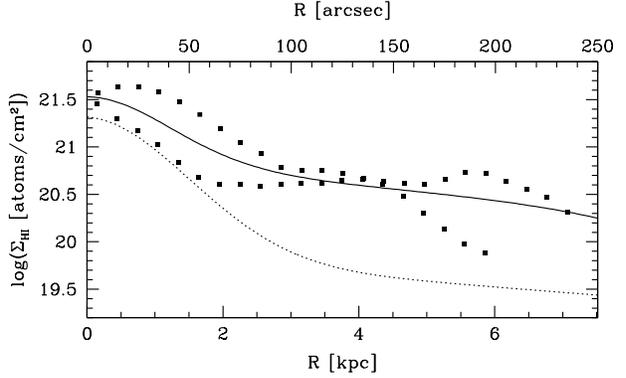,width=8.4cm}}}
\vspace{-3cm}
\caption{Observed radial profile (squares) of the \HI\ disk
extracted from the same 40 arcsec  wide slices of the NA and UN data cubes
used to derive the rotation curve (e.g.\ Fig~\ref{f:pv1_na}). The two
``strands'' of data represent receding and approaching sides.  The solid
lines show the model edge-on disk fit to the average profile, while the
broken line shows the corresponding face-on profile.\label{f:nhi}}
\end{figure}

The radial surface density profiles we use in Section \ref{ss:mm} to
model the rotation curve have been derived relative to the dynamical
centre\footnote{The density distribution of the dark matter halo,
is not greatly affected if the optical or \HI\ centres are used to
extract the profiles.}  As illustrated in Fig.~\ref{f:nhi} The face-on
\HI\ profile $\Sigma_{\rm HI}(R)$ was modelled from the run of
integrated \HI\ flux with position in the PV cuts used to extract the
rotation curve.  For the model we assumed that the PV cuts represent the
integrated flux of an edge-on disk.  For the intrinsic
$\log(\Sigma_{HI})$ profile we adopt a linear decline with a Gaussian
core (i.e.\ similar to the model used by M96).  This functional form is
adopted merely because it fits the data fairly well, it is not meant to
provide insight to the physics of the mass distribution.  An arbitrary
truncation radius of $R_{\rm max} = 300''$ is adopted in order to insure
a finite disk mass, and a constant of 0.08 dex is added to the fitted
$\log(\Sigma_{HI})$ profiles in order to recover the \HI\ flux outside
of the extraction strip.  The fitted \HI\ profile is given by:
\begin{equation} 
\log \Sigma_{\rm HI} = 19.86 - 0.00169 R + 1.46 \label{e:fcprof}
\exp(-[R/50.2]^2/2), 
\end{equation} 
where $R$ is in arcsec, and $\Sigma_{\rm HI}$ is the face-on \HI\ column
density in atoms cm$^{-2}$.

The displacement of the optical centre from the dynamical centre is
36 arcsec; a distance more than two optical scale lengths as measured in
Paper I.  Averaged over their orbits about this centre, the stars have a
significantly more diffuse distribution than given in Paper I.
Therefore, for the mass modelling we adopt an optical light profile that
is exponential with an effective radius equal to the projected offset
between the dynamical and optical centres. The total brightness is taken
to be $m_{B,0} = 12.92$ (Paper I).  This does not include the light of
NGC1705-1 because while it supplies significant luminosity it has a
relatively negligible mass (Paper I; Ho \&\ Filipenko,
1996\nocite{hf96}). This yields a profile with central surface
brightness $\mu_{0}(B) = 21.59$ and scale length $\alpha^{-1}_B = 21.6''
= 0.65$ kpc.

\subsection{Pressure support and beam smearing\label{ss:ps}}

Figure~\ref{f:rc} shows the radial profile of the \HI\ velocity
dispersion $\sigma_{\rm HI}$ in its bottom panel; $\sigma_{\rm HI} \ga
15$ \kms\ at all radii, with a central value of about 28 \kms. This is
significantly broader than is observed in quiescent spiral galaxies.
These typically have $\sigma_{\rm HI} \approx 6 - 10$ \kms\ with mildly
increasing values towards the centre (e.g.\ Shostak \&\ van der Kruit
1984; Dickey, Hanson \&\ Helou 1990)\nocite{sv84}\nocite{dhh90}. If
these are true measures of the turbulence in the ISM, then pressure
support may be significant since $\sigma_{\rm HI}/V_{\rm rot} \sim 1$.
However, $\sigma_{\rm HI}$ may also be an artifact of differing bulk
motions along the long sight line through the disk, or beam smearing of
an unresolved and steep velocity gradient.  The latter is particularly a
concern in the centre of the galaxy, since the flat part of the rotation
curve is reached in only $\sim 3$ beam diameters.

Since beam smearing and centralized pressure support have a similar
effect - to decrease the central velocity gradient - correcting for
one will have a similar effect as correcting for the other.  In
Fig.~\ref{f:rc} we show the effect of correcting the rotation curve
for ``minimum'' and ``maximum'' levels of pressure support, following
the method outlined by M96.  The assumed $\sigma_{\rm HI}$ curves for
the two cases are shown as dotted and long-dashed lines
respectively. The minimum case, a flat $\sigma_{\rm HI} = 6$ \kms\ at
all radii, assumes that the observed $\sigma_{\rm HI}$ profile is
dominated by projection effects, and that the face-on $\sigma$ is like
that seen in quiescent disk galaxies. The maximum case, shown with the
long dashed line, assumes that the observed $\sigma$ profile reflects
the real turbulent motions.  The top panel shows that the minimum
pressure support correction is negligible.  The maximum case, however,
makes a significant difference to the rotation curve; the turnover
radius moves inwards and the $V_c$ curve is significantly steeper in
the centre than $V_{\rm rot}$.  However there is little change of
$V_{\rm rot}$ at large $R$.

\subsection{Mass models\label{ss:mm}}

\begin{figure}
\centerline{\hbox{\psfig{figure=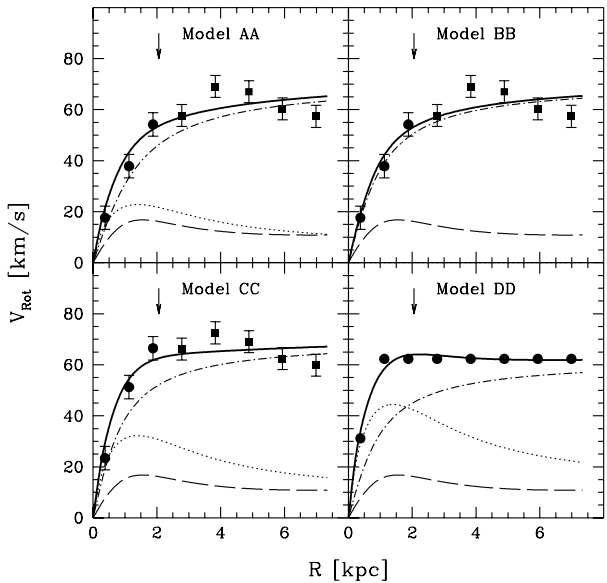,width=8.4cm}}}
\caption{Mass model fits to the derived rotation curve.  In each panel,
the full model is shown as a thick solid line.  The contributions of the
stellar, neutral ISM, and dark matter halo are shown with dotted,
dashed, and dot-dashed lines respectively.  The vertical arrows mark the
Holmberg radius as measured in Paper I.  The parameters for the
displayed mass models are listed in Table~\ref{t:vrfit}.\label{f:vrfit}}
\end{figure}

Mass model fits to the rotation curve are shown in Fig.~\ref{f:vrfit}.
The corresponding parameters of the models are listed in
Table~\ref{t:vrfit}.  The models shown consist of three components to
the mass distribution: (1) the stellar distribution which is given by
the projected luminosity profile (Section \ref{ss:prof}) scaled by \Mlbstar\
-- the mass to light ratio of the stars; (2) the neutral ISM
distribution which is set by the \HI\ profile scaled by a constant 1.33
to account for the Helium contribution; and (3) a dark matter (DM) halo.
The DM halo is taken to have a density distribution given by
\begin{equation} 
\rho = \frac{\rho_0}{1 + (R/R_c)^2} \label{e:mm}
\end{equation}
where the free parameters are the central density $\rho_0$ and the core
radius $R_c$. For this density
distribution the rotational velocity at large $R$, $V_{\infty}$, and halo
velocity dispersion $\sigma_0$ are given by \cite{lsv90}:
\begin{equation} 
V_{\infty}^2 = 4 \pi G \rho_0 R_c^2 = 4.9 \sigma_0^2. 
\end{equation}

\begin{table}
\caption{Mass models\label{t:vrfit}}
\begin{tabular}{lrrrr}
Model:                                             & AA   & BB   & CC   & DD \\ \hline
\underline{Fit parameters} : \\
\zsp ${\cal M}/L_B$ (solar units)     & 0.5    & 0.0    & 1.0    & 1.9   \\
\zsp $R_c$ (kpc)                      & 1.12   & 0.84   & 0.74   & 0.80  \\
\zsp $\rho_0~{\rm (\Msun pc^{-3})}$   & 0.076  & 0.130  & 0.163  & 0.111 \\
\zsp $\sigma_0$ (\kms)                & 23.9   & 23.6   & 23.2   & 20.7  \\
\zsp $V_\infty$ (\kms)                & 52.9   & 52.2   & 51.4   & 45.8  \\
\zsp rms (\kms)                       & 5.2    & 4.9    & 5.0    & 1.8   \\[0.5ex]
\underline{At $R_{\rm Ho} = 2.1$ kpc} : \\ 
\zsp $V_c$ (\kms)                                  & 53.7 & 53.4 & 62.6 & 64.0 \\
\zsp ${\cal M}_T ~(10^9\, \Msun)$                  & 1.30 & 1.34 & 1.74 & 1.72 \\
\zsp ${\cal M}_T/L_B$ (solar units)                & 3.2  & 3.3  & 4.3  & 4.2  \\
\zsp ${\cal M}_{\rm dark}/{\cal M}_{\rm Luminous}$ &  6.4 & 12.8 &  3.1 &  1.4 \\[0.5ex]
\underline{At last point, $R = 7.0$ kpc} : \\
\zsp $V_c$ (\kms)                                  & 64.9 & 65.1 & 67.0 & 61.9 \\
\zsp ${\cal M}_T ~(10^9\, \Msun)$                  & 6.81 & 6.86 & 7.23 & 6.14 \\
\zsp ${\cal M}_T/L_B$ (solar units)                & 16.8 & 16.9 & 17.9 & 15.1 \\
\zsp ${\cal M}_{\rm dark}/{\cal M}_{\rm Luminous}$ & 18.2 & 42.4 & 12.0
&  5.7 \\ \hline
\end{tabular}
\end{table}

The mass model fits employed the $\chi^2$ minimization code employed by
M96.  All models shown have fixed \Mlbstar. In most cases they are
maximum disk solutions where the maximum \Mlbstar\ value is determined
that will allow the innermost $V_{\rm rot}$ measurement to be be
modelled entirely by the stellar and ISM contributions.  The adopted
\Mlbstar\ is then set at 90\%\ of this value, reduced so that the DM
core is not hollow (cf.\ van Albada et al.\ 1985\nocite{vbbs85}).  We
prefer maximum disk models because there are not enough data to
meaningfully fit \Mlbstar\ (see below); the photometric profile only
covers the innermost two points of the rotation curve.  Furthermore,
these models minimize the contribution of the halo.  The fact that the
halo dominates in all fits shows that it is a necessary component of the
mass distribution.

In all models shown, the geometries of the stellar and neutral gas
distributions are assumed to be circularly symmetric razor thin disks, while
the DM halo is assumed to be spherically symmetric.  These standard
assumptions for modelling rotation curves, are of course highly idealised.
Considering that the disks in NGC~1705 may be rather thick, and/or warped, we
also constructed models with spherically symmetric distributions (having the
same enclosed ${\cal M}(R)$ relationship) for the gas and stars.  This
decreases both the maximum amplitude of the rotation curve of the two disk
components, and the radius at this maximum.  For maximum stellar disk models,
this results in a somewhat lower \Mlbstar, since the peak then occurs
somewhat closer to the innermost $V_{\rm rot}$ measurement, and consequently
a lower \Mlbstar\ is required to reach the same amplitude.  However the
assumptions about the geometry of the luminous components has only a minor
affect on the DM halo properties ($< 25$\%\ for $R_c$ and $\rho_0$; $< 0.2$
\kms\ for $\sigma_0$), since the halo is so dominant in all cases.

Model AA shows the maximum disk model for the observed rotation curve.
The derived $\Mlbstar = 0.5\, \Mlbsun$ is close to the $\Mlbstar = 0.6\,
\Mlbsun$ level estimated in Paper I from the HSB and LSB stellar
populations' colours.  Model BB is a fit to the same data assuming a
minimum stellar disk, i.e.\ $\Mlbstar = 0$, resulting in DM halo with a
core $\approx 30$\%\ smaller and denser by $\approx 60$\%.  When
\Mlbstar\ is a free parameter our minimization code also yields
$\Mlbstar = 0$, that is, no stellar component is required to fit these
data.  This is a common result when mass modelling rotation curves
\cite{dm97}, and indicates that the rotation curves do not often well
constrain the relative contribution of the stellar component even in the
core of galaxies - there is a degeneracy between \Mlbstar\ and $\rho_0$.
The minimum disk model yields the maximum DM $\rho_0$ allowed by these
data, or equivalently the total spherically averaged DM plus stellar
density within the constant density core of the galaxy.  A comparison of
$\rho_0$ in models AA and BB indicate that DM dominates even in the core
of NGC~1705.

Models AA and BB were fitted to the raw rotation curve derived in
Section \ref{ss:rc}.  However, as noted in Section \ref{ss:ps}, \Vrot\
in the inner most portion of the rotation curve may be underestimated
due to beam smearing and or pressure support.  Model CC is a maximum
disk fit to the rotation curve corrected for maximum pressure support
shown in Fig.~\ref{f:rc}.  A significantly higher \Mlbstar\ than in
model AA is needed to meet the maximum disk constraint.  Even so,
$\rho_0$ is almost twice that in model AA, and $R_c$ is smaller by 35\%.
Note that there is little effect on $\sigma_0$ (it actually goes down
slightly).  This is because it is essentially set by the flat part of
the rotation curve.

The maximum pressure support corrections may be insufficient to recover
the true \Vrot\ profile.  Beam smearing and warping of the
\HI\ disk may further lower \Vrot\ over its initial rise.  It
is impossible to sort out all these effects with the data in hand.
Instead, we construct an idealized rotation curve based on the shape of
normal galaxy rotation curves, and scaled to the properties of NGC~1705.
It features a linearly rising inner portion and a flat rotation curve
thereafter.  The turnover is set to occur where the stellar disk
rotation curve should peak, while the flat part at $\Vrot = 62.3$ \kms\
is set as the mean of the last five points in Table~\ref{t:rc}.  Model
DD is the maximum disk fit to this hypothetical rotation curve.  This
results in the highest \Mlbstar\ of our models.  Compared to model AA,
$R_c$ is smaller by 30\%\ and $\rho_0$ is larger by over 40\%, while
$\sigma_0$ is lower by 14\%.

The main conclusion from our mass models is that in all cases a dominant
DM halo is required.  Moreover, the central density $\rho_0 \sim 0.1\,
\Msun\, {\rm pc^{-3}}$ is rather large compared to other disk galaxies.
This result is discussed in detail in Section \ref{ss:ev}. 

\section{Discussion}\label{s:disc}

\subsection{Is NGC~1705 an interacting merging system?}

Our study shows that NGC~1705 has a lopsided and disturbed \HI\
appearance suggesting that an interaction or merger has triggered the
starburst in this system.  Although this is a very appealing scenario,
it is not clear what it is interacting with.  NGC~1705 inhabits a low
density environment.  There are no galaxies of comparable brightness or
size in the 15$'$ (27 kpc) field centred on NGC~1705 visible on the UK
Schmidt Telescope Survey.  We used NED\footnote{The NASA/IPAC
Extragalactic Database (NED) is operated by the Jet Propulsion
Laboratory, California Institute of Technology, under contract with the
National Aeronautics and Space Administration.} to search for possible
companions within 5$^\circ$ (0.54 Mpc in projection) and found no likely
companions.  Over this area there are only two galaxies as bright or
brighter than NGC~1705: NGC~1596 and NGC~1617.  These have significantly
higher $V_r$ (1510 and 1063 \kms\ respectively) indicating that they are
background galaxies.  The nearest catalogued group of galaxies is the
NGC~1672 group, with eight known member \cite{g93}, at $V_r = 1084$
\kms\ and a projected separation of $\approx 6.4^\circ \approx 0.7$ Mpc.

We searched our data for possible companions.  To do this we made a
special purpose NA data cube with a pixel size $15'' \times 15''$ in the
spatial direction (yielding a synthesized beam width of 44$'$) and 13.2
\kms\ in velocity, and large enough to easily encompass the ATCA primary
beam FWHM $= 34'$ ($R = 31$ kpc). The useful velocity range of this cube
is $V_r = 430 - 1220$ \kms (i.e.\ $\approx V_r({\rm
NGC1705})_{-200}^{+600}$ \kms). We examined the data cube in a variety
of ways, but found no obvious companions.  The estimated $S/N = 5$
detection limit corresponds to $\MHI \sim 10^6\, \Msun$ for a point
source of narrow velocity width ($W_{50} = 26$ \kms) located at the
field centre.  However, the area we surveyed is small compared to that
used by Taylor et al.\ \shortcite{tbgs95} study.  Of the \HII\ galaxy
companions, $\approx 60$\%\ have projected separations $> 31$ kpc.
Hence there may be a companion beyond the area we surveyed.

Perhaps the \HI\ spur is actually a companion galaxy.  If so it must
have a low stellar content since the image of NGC~1705 in the ESO-LV
catalogue \cite{lv89} shows no optical counterpart to the spur, down to
its detection limit of $\mu_B \approx 24\, {\rm mag\, arcsec^{-1}}$.  An
examination of the NA data cube reveals that the spur is attached to the
main body of the system both spatially, and in $V_r$, suggesting that if
it is a companion, it is strongly interacting or merging with NGC~1705.
However, NGC~1705's outer optical isophotes are round and regular,
suggesting a relatively peaceful environment. This is in contrast to the
best studied merger-BCD: IIZw40 (Brinks \&\ Klein, 1988; van Zee,
Skillman, \&\ Salzer 1998)\nocite{bk88}\nocite{vss98} which has a highly
disturbed ``$\times$'' structure even at low surface brightness levels
(Marlowe et al.\ 1997\nocite{mmhs97}; Telles, Melnick, \&\ Terlevich
1997\nocite{tmt97}). We conclude that there is no evidence that the spur
is a companion galaxy and hence we have not found any convincing
external trigger to NGC~1705's starburst.

\subsection{Evidence for a galactic wind blow out}\label{ss:blow}

In the standard theory of galactic winds \cite{ham90} the wind
expansion is powered by a very hot, $T = 10^7 - 10^8$ K,
over-pressurized bubble of thermalized supernovae remnants and stellar
winds, which are abundantly produced in a starburst.  As it expands the
bubble sweeps up the ambient ISM into a shell like structure.  Once it
expands to one or two disk scale heights it will accelerate out along
the minor axis, resulting in shell fragmentation.  The enclosed hot gas
will then ``blow out'', escaping into the halo.  We propose that the
\HI\ spur results from such a galactic wind blow out.  The evidence is
as follows.

Firstly, the orientation of the spur  is consistent with the optical
outflow.  The position angle of the spur $\phi \approx -10\degr$ relative
to the optical centre, while the \Halpha\ outflow
axis $\phi \approx -20\degr$ (Paper I).  Furthermore, $V_r$ increases gradually
with distance from the optical centre, the same sense as the optical
velocity gradient on the NNW side of the outflow. 

Secondly, the one-sided nature of the spur is predicted by models of
galactic winds.  If the energy source is displaced off of the disk
plane, the blow out will occur only on the side to which the source is
displaced (Mac Low, McCray, Norman 1989)\nocite{mmn89}.  Indeed, the
spur is displaced on the same side of the \HI\ disk plane as the optical
centroid.

Thirdly, the ISM is very porous towards the optical centre of
NGC~1705. Heckman \&\ Leitherer \shortcite{hl97} estimate $N_{\rm HI} =
1.5 \times 10^{20}\, {\rm cm^{-2}}$ towards NGC1705-1 from the \Lya\
absorption profile obtained with the Hubble Space Telescope, over ten
times lower than what we find $N_{\rm HI} = 2\times 10^{21}\, {\rm
cm^{-2}}$ averaged over the UN beam.  The observed weakness, and
porosity in the ISM suggests that the hot bubble has already punched
through the ambient ISM.

Finally, the scale size of the spur, 4.5 kpc, agrees with expectations.
Once the blow out occurs, the hot gas will expand at thermal velocities
and adiabatically cool.  The thermal velocity for $T = 10^7$ K is 280
\kms\ for an H atom.  Thus over a 10 Myr time-scale (the optical
expansion time-scale), gas from the blow-out can travel nearly 3 kpc, 3/4
the length of the spur.

One problem with this scenario is that it implies that the outflow
vector must be inclined within only a few degrees of the plane of the
sky, otherwise the spur would have a much larger $V_r$ difference with
respect to $V_{\rm sys}$, (i.e.\ even larger than $i \approx 68\degr$
derived in Paper I for the \Halpha\ outflow).  However, NGC1705-1 is but
the brightest cluster formed within the young central stellar population
population.  So perhaps earlier star formation events expelled the spur
at lower velocities.  At a minimum, the velocity of the spur is $\sim
30$ \kms, the $V_r$ difference with respect to $V_{\rm sys}$, yielding a
maximum expansion time-scale of about 100 Myr.

What will become of the galactic wind? Already some of it has travelled
$\sim 4.5$ kpc - well into the halo. The escape speed $V_e$ at radius
$R$ within an isothermal sphere truncated at $R_{\rm max}$ is given by
\begin{equation}
V_e(R) = \sqrt{2}V_c\left[1+\ln\left(\frac{R_{\rm max}}{R}\right)\right]^{1/2}
\end{equation}
where $V_c$ is the circular velocity.  Thus if the spur is travelling at
280 \kms\ at $R = 3$ kpc, the DM halo would have to extend out to $\sim
30$ Mpc to contain the outflow.  This is implausibly large, hence the
spur will escape.  If $R_{\rm max}$ is as low as 7 kpc (the limits of
the \HI\ data) the spur need only be travelling at 120 \kms\ for it to
escape.  The neutral gas (H + He) mass of the spur is $1.7 \times 10^7$
\Msun, about 8\%\ of the total gas mass.  If it results from an outflow,
then the range of plausible time scales ($\sim 10 - 100$ Myr) imply mass
loss rates of $\sim 0.17 - 1.7\, \Msun\, {\rm yr^{-1}}$, into the DM
dominated halo if not out of the system entirely.  This is roughly equal
to or up to an order of magnitude larger than the {\em current\/} total
star formation rate of $0.13\, \Msun\, {\rm yr^{-1}}$ as derived from
the \Halpha\ flux (Paper I) and the conversion factor of Kennicutt
\shortcite{k83}.

\subsection{Disk stability and star formation}

The link between star formation and the gravitational stability of the
ISM in a disk has been noted by several authors over the past few
decades. (e.g.\ Quirk, 1972\nocite{q72}; Kennicutt 1989\nocite{k89}).
If the disk is sufficiently cool and dense its thermal pressure and
centripetal acceleration can not support itself against self gravity and
the disk will fragment, presumably leading to efficient star formation.
The gravitational stability of a gaseous disk is given by the Toomre
\shortcite{t64} $Q$ parameter:
\begin{equation}
Q \equiv \frac{\sigma_g \kappa}{\pi G \Sigma_g}\label{e:q}
\end{equation}
where $\sigma_g$ is the gas velocity dispersion, $\Sigma_g$ is the gas disk
surface density and 
\begin{equation}
\kappa = \left(R\frac{d\Omega^2}{dR} + 4\Omega^2\right)^{1/2}\label{e:k}
\end{equation}
is the epicyclic frequency for angular frequency $\Omega = V_{\rm
rot}/R$.  High values of $Q$ mean the disk is more stable while if $Q
\la 1$ then the disk should fragment.  Kennicutt \shortcite{k89}
empirically found that in normal disk galaxies, high mass star formation
occurs in regions having $Q \la 1.6$.  The same threshold also is able
to predict the location of star formation in low surface brightness
galaxies \cite{vsbm93} and BCD/\HII\ galaxies \cite{tbps94}.

\begin{figure}
\centerline{\hbox{\psfig{figure=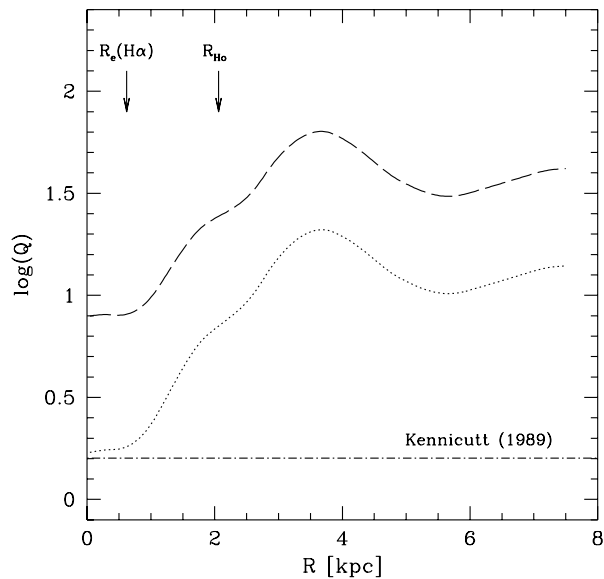,width=8.4cm}}}
\caption{Radial variation of Toomre's disk stability
parameter $Q$.  The solid and dotted lines correspond to maximum and
minimum pressure support (see Fig.~\ref{f:rc}). The dot-dashed line
marks the threshold $Q$ value for efficient star formation as derived by
Kennicutt (1989).  The arrows mark the \Halpha\ effective radius and the
Holmberg radius.\label{f:qprof}}
\end{figure}

Figure~\ref{f:qprof} shows NGC~1705's azimuthally averaged $Q$ profile
for the minimum and maximum $\sigma_{\rm HI}$ curves of Fig.~\ref{f:rc}.
Here $V_{\rm rot}$ is interpolated smoothly (spline fit) through the
points given in Table~\ref{t:rc}.  The $\Sigma_g$ profile is taken as
the neutral gas surface density profile fits given in
eq.~\ref{e:fcprof}.  Two fiducial radii are indicated in this plot:
$R_{\rm Ho}$ which encompasses most of the stellar light, and the
\Halpha\ effective radius \cite{mmhs97} which encompasses most
or all of the high mass star formation.  We see that the $Q$ profiles
has a sharp down-turn corresponding to the optical extent of the galaxy
and that high mass star formation is limited to the region where the $Q$
profile bottoms out.  Conversely, the outer \HI\ disk (beyond $R_{\rm
Ho}$) is relatively stable with $Q$ a factor of 5-10 higher than in the
centre of the galaxy.

When the minimum $\sigma_{\rm HI}$ profile is adopted the central values
of $Q$ in NGC~1705 are close to Kennicutt's threshold value.
Considering the limited resolution of the data and the neglect of
molecular gas and elements heavier than helium, we may expect that the
disk is even less stable in the centre.  The maximum $\sigma_{\rm HI}$
version of the $Q$ profile is significantly above Kennicutt's threshold
at all radii.  However, it should be pointed out that Kennicutt
\shortcite{k89} as well as van der Hulst et al.\ \shortcite{vsbm93} and
Taylor et al.\ \shortcite{tbps94} {\em assumed\/} $\sigma_g = 6$ \kms\
at all radii.  Thus the threshold $Q$ has yet to be calibrated with true
measured $\sigma_g$ profiles.  We conclude that NGC~1705 is like most
other galaxies in that star formation occurs where the disk is least
stable.

\subsection{Dark matter halos and dwarf galaxy morphology\label{ss:ev}}

\begin{figure}
\centerline{\hbox{\psfig{figure=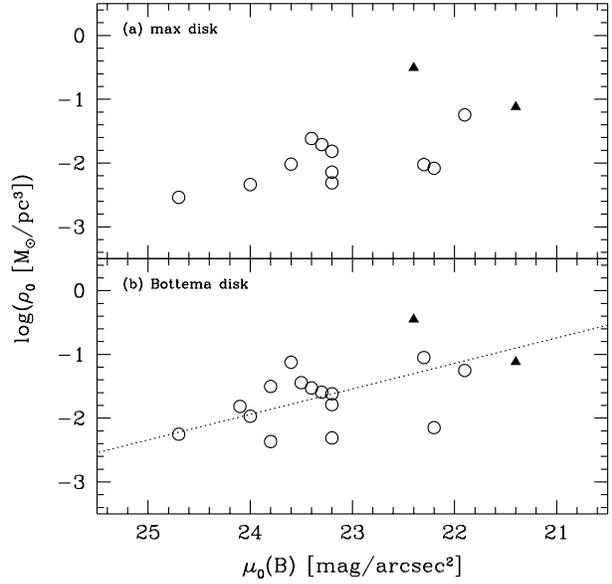,width=8.4cm}}}
\caption{DM Halo core density $\rho_0$ compared to the
optical ($B$ band) extrapolated central surface brightness $\mu_0(B)$.
Circles are from the models of de Blok \&\ McGaugh (1997) and only
include galaxies with $M_B \geq\ -18$ in order to isolate dwarfs.
Triangles represent the two mass modelled BCDs NGC~1705 (on the right;
this study) and NGC~2915 (on the left; from M96 for $H_0 = 75\, {\rm
km\, s^{-1}\, Mpc^{-1}}$).  The top panel shows $\rho_0$ from maximum
disk mass models, while the lower panel shows $\rho_0$ when \Mlbstar\ is
set by the colour of the galaxy - the Bottema disk solution.  The
quantities plotted are correlated with correlation coefficients of
$-0.66$, $-0.53$ for the top and bottom panels respectively.  The dotted
line in the lower panel shows a least squares fit to the data with the
slope set to --0.4 dex per mag arcsec$^{-2}$ as expected from the disk
stability arguments given in Section \ref{ss:ev}.  Allowing the slope to vary
yields slopes of --0.41 and --0.32 dex per mag arcsec$^{-2}$ and rms
residuals of 0.41 and 0.42 dex for the top and botom panels
respectively.\label{f:halocor}}
\end{figure}

One important difference between BCDs and dI galaxies is in the central
densities $\rho_0$ of their DM halos, as shown in Fig.~\ref{f:halocor}.
The majority of the data comes from the uniform mass model fitting of de
Blok \&\ McGaugh \shortcite{dm97}.  Since we are concerned with dwarf galaxy
properties we limit their sample to systems with $M_B > -18$.  These
galaxies mostly have Sm or later morphological types (i.e.\ dI
galaxies), although there are a few Sd galaxies particularly at
$\mu_0(B) \approx 22\, {\rm mag\, arcsec^{-2}}$.  The two BCDs are
NGC~1705, from this study, and NGC~2915 from M96.  Both panels plot
$\rho_0$ of the DM halo against the optical $B$ band central surface
brightness $\mu_0(B)$.  For NGC~1705 we adopt $\rho_0$ from model AA,
since this gives the most conservative results.

The top panel shows the results of maximum disk models.  Since maximum
disk models correspond to minimum halo models, the $\rho_0$ plotted in
Fig.~\ref{f:halocor}a should be considered lower limits.  In order to
break the degeneracy between \Mlbstar\ and $\rho_0$ de Blok \&\ McGaugh
employ ``Bottema disk'' model fits; their results are plotted in the
bottom panel of Fig.~\ref{f:halocor}.  For these \Mlbstar\ is set by the
$(B-V)$ colour of the galaxy, rather than the rotation curve.  Following
their prescription and the photometry in Paper I we obtain $\Mlbstar =
0.8$ \Mlbsun\ for NGC~1705. This value is higher than the maximum disk
\Mlbstar, and following de Blok \&\ McGaugh we adopt the maximum disk
\Mlbstar\ results (model AA) for the Bottema disk solution.  Since
NGC~2915 was not modelled with a Bottema disk by M96, we estimate the
Bottema disk results here.  From the prescription of de Blok \&\ McGaugh
and the the photometry of Longo \&\ de Vaucouleurs \shortcite{ld83} we
obtain a Bottema disk $\Mlbstar = 1.0$ \Mlbsun\ for NGC~2915.  This is
slightly less than the maximum disk \Mlbstar\ in M96's model D. By
interpolation between their models A and D and correcting to $H_0 =
75\, {\rm km\, s^{-1}\, Mpc^{-1}}$ we estimate $\rho_0 = 0.36\, \Msun\,
{\rm pc^{-3}}$ for the Bottema disk solution.

Figure~\ref{f:halocor} shows that the two BCDs have some of the highest
$\rho_0$ of any of the dwarf galaxies in the sample.  Furthermore there
is a weak but noticeable correlation between $\log(\rho_0)$ and
$\mu_0(B)$, with higher surface brightness disks corresponding to higher
$\rho_0$ halos.  These results hold for both maximum disk and Bottema
disk solutions. This result (for Bottema disks) was noted first by de
Blok \&\ McGaugh \shortcite{dm97}, and extends to high luminosity disks.  Here we
show that the correlation also includes blue compact dwarf galaxies.
The rms scatter in $\log(\rho_0)$ about simple linear fits to the data
is $\approx 0.4$ dex (see Fig~\ref{f:halocor} caption).  A large part of
this may be due to measurement uncertainties; as noted the pressure
support or beam smearing corrections can change $\log(\rho_0)$ by 0.3
dex, while distance errors can easily translate into a 0.2 dex
uncertainty in $\log(\rho_0)$ for galaxies with $D \leq 5$ Mpc.

The fact that the two BCDs are on the right side of the diagram while
most of the dIs are on the left side is not a coincidence.  It has been
noted by several authors (Papaderos et al.\ 1996\nocite{plft96}; Telles
\&\ Terlevich, 1997\nocite{tt97}; Marlowe, Meurer, \&\ Heckman
1998\nocite{mmh98}) that the enveloping exponential profile of BCDs have
brighter $\mu_0(B)$ than dIs by typically $\sim$2.5 mag arcsec$^{-2}$
\cite{mmh98}.  Note that the $\mu_0(B)$ plotted here excludes any
central excess of light above the usual exponential profile seen at
large radii.  Such an excess of light is commonly found in BCD galaxies
and usually is identified as the starburst \cite{mmh98}.  They are also
common in dI galaxies, but at a lower intensity \cite{pt96}, and hence
are not classified as starbursts.

What is the explanation of the differences between BCDs and dIs apparent
in Fig.~\ref{f:halocor}?  We suggest that DM halos regulate the
morphology of dwarf galaxies via the stability of their inner disks.
Both types of dwarfs are usually DM dominated at nearly all radii, and
typically have solid body rotation curves over their optical extents.
Let us assume that star formation occurs at the same $Q$ value in all
types of dwarf galaxies, and moreover that $Q$ is regulated to maintain
this level, at least over the optical extent of the galaxy (cf.\
Ferguson, 1997\nocite{f97}; Kennicutt, 1989\nocite{k89}).  For a given
$Q$, in the solid body portion of the rotation curve $\Sigma_g \propto
\kappa = 2\Omega \propto \rho_0^{1/2}$ (from eq.~\ref{e:q} and
\ref{e:k}).  Kennicutt \shortcite{k98} finds that the the global star
formation rate per unit area in disk galaxies $\propto \Sigma_g/t_{\rm
dyn}$ where $t_{\rm dyn} \approx (G\rho)^{-1/2}$ is the dynamical time
scale.  Combining this with our constant $Q$ assumption leads to
\begin{equation}
\frac{\rm star~formation~rate}{\rm area} \propto \rho_0 \propto
\Omega^2.\label{e:sbcor}
\end{equation}
The optical luminosity of both dIs and BCDs are usually dominated by
young blue stars, so the star formation rate per area is proportional to
surface brightness.  Hence the $\sim 1$ dex difference in $\rho_0$
between dIs and BCDs should correspond to a factor of ten difference in
linear surface brightness, or 2.5 mag arcsec$^{-2}$.  The best fit line
with this slope is shown in Fig.~\ref{f:halocor}.  We see that this
scenario can explain the basic trend seen in the data.  We emphasize
that this model explains the differences in the exponential envelopes
$\mu_0(B)$ of gas rich dwarfs and does not address the starburst nature
of BCDs.  As shown by Marlowe et al.\ \shortcite{mmh98} the starburst
typically is at best a modest enhancement to the luminosity of a BCD and
contributes only a few percent to the stellar mass of the host.

While the scatter in Fig.~\ref{f:halocor} allows for some evolution in
surface brightness, our scenario implies that an extreme dI galaxy can
not evolve into an extreme BCD, or visa-versa.  This would involve a factor of 10
change in $\rho_0$, or expansion/contraction of the halo by a factor of
$10^{1/3}$.  This is very difficult to do if DM is non-dissipitive, and
since even the cores are DM dominated, a baryonic (luminous matter)
collapse or blowout would be insufficient to drag the DM with it.

\section{Conclusions}\label{s:conc}

We have examined the \HI\ distribution and dynamics of the windy blue
compact dwarf (BCD) galaxy NGC~1705.  While a rotating disk dominates
the \HI\ distribution, the \HI\ distribution appears somewhat disturbed.
The optical centre and \HI\ peak are both offset from the \HI\ dynamical
centre radially and off of the \HI\ plane.  The total projected offset
amounts to about 1.1 kpc.  In addition there is an extraplanar \HI\ spur
extending at least 4.5 kpc into the halo which accounts for $\sim 8$\%\
of the total \HI\ mass.  We argue that it is likely to have a galactic
wind origin, perhaps being the adiabatically cooled hot bubble that
formerly powered NGC~1705's spectacular \Halpha\ outflow.  Higher
resolution observations could shed more light on this issue.

NGC~1705's rotation curve is similar in form to those of other gas rich
dwarf galaxies, having a linearly rising (solid body) inner portion and
then turning over to become approximately flat ($v_\infty = 62$
\kms).  Mass model fitting  shows that the dark matter
(DM) halo is dominant at nearly all radii - even into the solid body
core of the galaxy.  The models yield a DM halo central density $\rho_0
\approx 0.1\, \Msun\, {\rm pc^{-3}}$, which is about a factor of ten
times higher than typically found in dwarf irregular (dI) galaxies, but
similar to that found in NGC~2915 the only other mass-modelled BCD.

This study has provided useful insights into the dynamical evolution of
dwarf galaxies.  We can now address the four questions presented in the
introduction. 

Firstly, if NGC~1705's \HI\ spur is wind ejecta, then $\sim 2 \times
10^7\, \Msun$ of neutral ISM is being expelled at least into the halo if
not out of the system entirely.  This is occurring over a time-scale
between roughly 10 Myr and 100 Myr; yielding mass loss rates between 0.2
and 2 \Msun\ ${\rm yr^{-1}}$.  This is a significant mass loss event, up
to an order of magnitude larger than the current star formation rate.
However, we emphasize that the vast majority of the neutral ISM remains
in a disk structure.

Secondly, it is not clear from this study what triggered the starburst
in NGC~1705.  No clear external trigger was identified in our data, nor
in catalogue searches.  A secular origin for the starburst may also be
tenable: star formation in this galaxy occurs where the disk is least
stable to self gravitation - just as is seen in normal disk galaxies.

Thirdly, the large difference in DM halo central densities between BCDs
and dIs strongly indicates that there is little evolution between the
two types.  This is consistent with the differences found between the
optical structure of dIs and the host galaxies of BCDs (i.e.\ excluding
the starburst; Marlowe et al.\ 1998).  

Finally, we argue that dominant DM halos can regulate star formation in
dwarf galaxies.  This is done by setting the critical surface density
for self gravitation (and hence star formation) of the embedded baryonic
disk.  In this scenario surface brightness is correlated with $\rho_0$
so that BCDs should have stellar disk surface brightnesses $\sim$ 2.5
mag arcsec$^{-2}$ brighter than dIs, as is typically observed.

This paper benefited from discussions with Claude Carignan, Ken Freeman,
John Reynolds, Stacy McGaugh, and Chris Mihos. We thank the staff at
Paul Wild Observatory, Narrabri, in particular Robin Wark and Mike
Kesteven, for their support of our observing runs.  Claude Carignan
kindly provided some of the software used in this project.  GRM thanks
the director of ATNF, Ron Ekers, for the hospitality shown during three
visits to the Marsfield site. GRM gratefully acknowledges receipt of
travel funds provided by the American Astronomical Soceity and the
Center for Astrophysical Sciences of The Johns Hopkins University, which
made this collaboration tenable from.  Examination of our data cubes
benefitted from the KARMA packaged developed at the ATNF by R.\ Gooch.
Literature searches were performed using NED, the NASA/IPAC
Extragalactic Database, a facility operated by the Jet Propulsion
Laboratory, Caltech, under contract with the National Aeronautics and
Space Administration.

\end{document}